**Motion of a spherical solid particle in Couette flow: exact solution vs. homotopy perturbation approximation with and without Padé approximants**

Tarek M. A. El-Mistikawy

Department of Engineering Mathematics and Physics, Cairo University, Giza 12211, Egypt

**Abstract**

The motion of a spherical solid particle in plane Couette flow is governed by a linear problem that has a simple exact solution. As such, there is no need for an approximate analytical representation of the solution; specially when it is tedious, complicated, and requires hairy terms to give accurate results only at small or moderate values of the time.

**Introduction**

The motion of a spherical solid particle in plane Couette flow is governed by the following linear problem

$$\ddot{x} - A\dot{y} + B(\dot{x} - \alpha y) = 0 \tag{1}$$

$$\ddot{y} + B\dot{y} + (C + A)(\dot{x} - \alpha y) = 0 \tag{2}$$

$$x(0) = 0, \dot{x}(0) = u \tag{3}$$

$$y(0) = 0, \dot{y}(0) = v \tag{4}$$

where dots denote differentiation with respect to time $t$.

This problem was first formulated by Vander Werff [1], who also obtained its closed form solution when $u$=0.

The solution when $u$≠0 can be easily obtained using Laplace transform. It is

$$x = \beta\alpha t + \gamma - \gamma e^{-Bt}\cosh\lambda t + (u - \beta\alpha - B\gamma)e^{-Bt}\frac{\sinh\lambda t}{\lambda} \tag{5}$$

$$y = \beta - \beta e^{-Bt}\cosh\lambda t + (v - B\beta)e^{-Bt}\frac{\sinh\lambda t}{\lambda} \tag{6}$$

where

$$\overline{C} = C + A, \lambda = \sqrt{\overline{C}(\alpha - A)} \neq B, \beta = \frac{vB - u\overline{C}}{B^2 - \lambda^2}, \gamma = \frac{(uB + vA) - 2B\beta\alpha}{B^2 - \lambda^2} \tag{7}$$

For $\lambda$=$B$, the solution becomes

$$x = \frac{(vB - u\overline{C})\alpha}{4B} t^2 + \frac{2B(uB + vA) - (vB - u\overline{C})\alpha}{4B^2} t + \frac{2B(uB - vA) + (vB - u\overline{C})\alpha}{8B^3} (1 - e^{-2Bt}) \tag{5'}$$

$$y = \frac{(vB - u\overline{C})}{2B} t + \frac{vB + u\overline{C}}{4B^2} (1 - e^{-2Bt}) \tag{6'}$$

Jalaal et al. [2] handled the problem, for the special case $A$=$B$=$C$=$\alpha$=$u$=$v$=1, using the homotopy perturbation method (HPM). They stated 5 terms of the expansions for $x$ and $y$, which combine to give

$$\begin{aligned} x &= (-e^{-t} + 1) + (e^{-t} + t - 1) + (-2te^{-t} - 5e^{-t} + \tfrac{1}{2}t^2 - 3t + 5) \\ &+ (t^2e^{-t} + 8te^{-t} + 19e^{-t} + \tfrac{1}{6}t^3 - \tfrac{5}{2}t^2 + 11t - 19) \\ &+ (-\tfrac{1}{3}t^3e^{-t} - 6t^2e^{-t} - 36te^{-t} - 81e^{-t} + \tfrac{1}{24}t^4 - \tfrac{7}{6}t^3 + \tfrac{21}{2}t^2 - 45t + 81) \\ &+ \cdots \end{aligned} \tag{8a}$$

$$\begin{aligned} y &= (-e^{-t} + 1) + (2te^{-t} + 3e^{-t} + t - 3) + (-t^2e^{-t} - 6te^{-t} - 11e^{-t} + \tfrac{1}{2}t^2 - 5t + 11) \\ &+ (\tfrac{1}{3}t^3e^{-t} + 5t^2e^{-t} + 24te^{-t} + 45e^{-t} + \tfrac{1}{6}t^3 - \tfrac{7}{2}t^2 + 21t - 45) \\ &+ (-\tfrac{1}{12}t^4e^{-t} - \tfrac{7}{3}t^3e^{-t} - 22t^2e^{-t} - 100te^{-t} - 191e^{-t} + \tfrac{1}{24}t^4 - \tfrac{3}{2}t^3 + \tfrac{35}{2}t^2 - 91t + 191) \\ &+ \cdots \end{aligned} \tag{8b}$$

They also presented results for up to 12 terms with reasonable accuracy achieved for $t$ up to 6, at most; as their Fig. 1 indicates.

Hamidi et al. [3] extended the accuracy of this homotopy perturbation solution to higher values of $t$ by representing more terms by Padé approximants; as their Figs. 5 and 6 indicate. They gave the [8/8] Padé approximant for the $x$-velocity component and the [10/10] Padé approximant for the $y$-velocity component as follows.

$$V_x = \frac{\begin{array}{l} 1+\frac{80630168}{150869313}t-\frac{94677997}{258633108}t^2+\frac{130919939}{1508693130}t^3 \\ -\frac{1097649283}{94142451312}t^4+\frac{352444733}{353034192420}t^5-\frac{573969527}{10355669644320}t^6 \\ +\frac{29057689}{15533504466480}t^7-\frac{213287989}{7117169319187200}t^8 \end{array}}{\begin{array}{l} 1+\frac{80630168}{150869313}t+\frac{34638557}{258633108}t^2+\frac{10391023}{502897710}t^3 \\ +\frac{203299561}{94142451312}t^4+\frac{55668097}{353034192420}t^5+\frac{27350129}{3451889881440}t^6 \\ +\frac{786311}{3106700893296}t^7+\frac{24009973}{6022220193158400}t^8 \end{array}} \quad \text{(9a)}$$

$$V_y = \frac{\begin{array}{l} 1-\frac{617044058999277705}{383360519991315667}t+\frac{2881320393477961463\,3}{2913539951933999069\,2}t^2 \\ -\frac{70652029874030657113\,7}{148590537548633952529\,2}t^3+\frac{1588250719693734831\,59}{148590537548633952529\,2}t^4 \\ -\frac{866056762902097291\,1}{782055460782283960680}t^5+\frac{16420817408783354039}{713234580233442972140\,16}t^6 \\ +\frac{871887087316144450\,57}{1248160515408525201245280}t^7-\frac{1062628804263470249167}{1298086936024866209295091\,20}t^8 \\ +\frac{569687662164414487\,97}{146034780302797448545697760}t^9-\frac{2600485162883807728\,73}{3671731619041764420577543680\,0}t^{10} \end{array}}{\begin{array}{l} 1+\frac{149676980983353629}{383360519991315667}t+\frac{786100576523938020\,3}{2913539951933999069\,2}t^2 \\ +\frac{215683905534451894091}{148590537548633952529\,2}t^3+\frac{22009213655154354015}{4953017918287798417\,64}t^4 \\ +\frac{42598825941687607239}{495301791828779841764\,0}t^5+\frac{404907655928398513021}{356617290116721486070080}t^6 \\ +\frac{43862642271172589551}{4160535051361750670817\,60}t^7+\frac{883706266259083441033}{1298086936024866209295091\,20}t^8 \\ +\frac{16590820145764536325}{5841391212111897941827910\,4}t^9+\frac{307482708619550706343}{514042426665847018880856115\,20}t^{10} \end{array}} \quad \text{(9b)}$$

The simple exact solution given by Eqs. (5)-(7) reduces for this special case {using $\lim_{\lambda\to 0}\frac{\sinh \lambda t}{\lambda}=t$ } to

$$x = 2-2e^{-t}-te^{-t},\ y = te^{-t} \quad \text{(10a,b)}$$

so that

$$V_x = (1+t)e^{-t},\ V_y = (1-t)e^{-t} \qquad (11a,b)$$

Comparing Expressions (8) to (10) and (9) to (11) one realizes how futile it is to handle this particular problem with an approximate analytical approach.